\documentclass[fleqn,12pt,twoside]{article}
\usepackage{espcrc1}

\usepackage{graphicx,epsfig}
\usepackage[figuresright]{rotating}

\newcommand{\AmS}{{\protect\the\textfont2
  A\kern-.1667em\lower.5ex\hbox{M}\kern-.125emS}}

\title{Relativistic Nuclear Collisions and the QCD Phase Boundary
}
 
\author{P. Braun-Munzinger\address{GSI and Darmstadt University of Technology,
        Planckstr. 1,\\  
        64291 Darmstadt, Germany}}
       
\begin{document}

\maketitle
\begin{abstract}
In nucleus-nucleus collisions at ultra-relativistic energies matter is
formed with initial energy density significantly exceeding the
critical energy density for the transition from hadronic to
partonic matter. We will review the experimental evidence for this new
form of matter - the Quark-Gluon Plasma - from recent experiments at
the SPS and RHIC with emphasis on collective behavior, thermalization,
and its opacity for fast partons. We will further show that one can
determine from the data a fundamental QCD parameter, the critical
temperature for the QCD phase transition.
\end{abstract}

\section{Introduction and historical remarks}
About 40 years ago Hagedorn \cite{hagedorn} realized that the then growing
evidence for an exponentially increasing mass spectrum of hadronic resonance
states would lead to divergencies in the energy density (and other
thermodynamic quantities) at the (Hagedorn) temperature T$_H \approx$
m$_{\pi}$. A few years later, after the discovery of asymptotic freedom in QCD
\cite{wilczek,politzer} it was realized by Collins and Perry \cite{collins}
and independently by Cabibbo and Parisi \cite{cabibbo} that a hadronic system
at sufficiently high density or temperature should convert into a new state of
matter, commonly called quark-gluon plasma (QGP), at a critical temperature
T$_c$.  In fact, the first phase diagram of nuclear/hadronic matter appears in
\cite{cabibbo}. Although T$_c$ and T$_H$ were not immediately related at the
time \footnote{See \cite{hagedorn1} for an interesting discussion on this
subject.}, it is now clear that both developments provided the starting point
for this growing field of research.  About 8 years later Baym \cite{baym} drew
up the phase diagram with many of the features discussed presently
\footnote{For a discussion of the events leading to the construction of the
  RHIC accelerator   see \cite{baym}.}.

\begin{figure}[hbt]
\begin{center}
\vspace{-1.0cm}
\includegraphics[width=10.0cm]{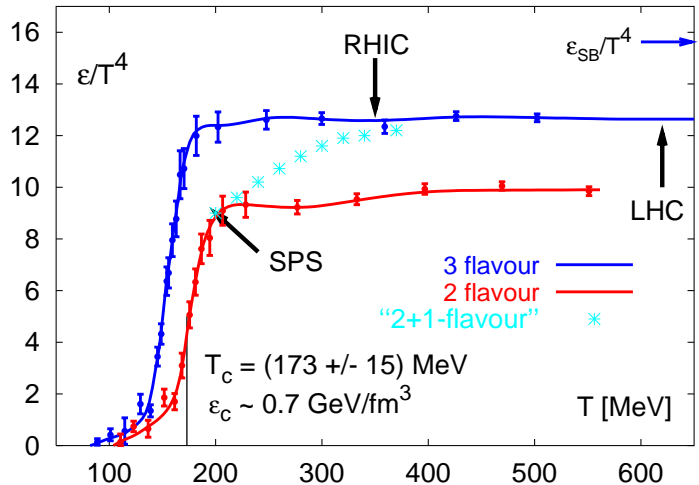}
\vspace{-.80cm}
\caption{Temperature dependence of energy density  in lattice calculations
  with 2 light and 1 heavier quark flavor \cite{karsch}}
\label{lattice}
\end{center}
\end{figure}

The physics of the transition from hadronic matter to QGP is now much better
understood, at least for vanishing net baryon density: solving QCD on a
discrete lattice has led to the results summarized in Fig.~\ref{lattice}. Near
a critical temperature of T$_c$ = 175 $\pm 15$ MeV the energy density (and
other thermodynamic quantities) exhibit a strong increase, signalling the
transition to a deconfined state consisting of quarks and gluons. The critical
energy density $\epsilon(\rm{T}_c$) is 700 $\pm$ 200 MeV/fm$^3$, about 5 times
the energy density in a large nucleus; whether the transition is a true phase
transition in the thermodynamic sense or rather a rapid cross-over is
currently much discussed \footnote{A cross-over seems more likely, see, e.g.,
\cite{karsch1}.} but for most of the experimental observables discussed below
not relevant.

\section{Hadron production and the QCD phase boundary}
A vigorous experimental program has been mounted over the past 20 years, first
at the BNL-AGS and CERN SPS accelerators and since the year 2000 at RHIC, to
collide atomic nuclei at ultra-relativistic energies with the aim to produce
the QGP and to study its properties. An impressive database has been assembled
in the center of mass energy range 4 $< \sqrt{s}_{nn} <$ 200 GeV. The current
review is not meant to be comprehensive, but rather focusses on collective
behavior of the matter produced, on its opacity to fast quarks, and on our
current knowledge of the phase boundary between QGP and hadronic matter. Much
more information on the experimental database as well as on various
theoretical descriptions can be found in the proceedings of recent Quark
Matter conferences \cite{nantes,oakland}.

\begin{figure}[hbt]
\begin{center}
\vspace{-3.0cm}
\includegraphics[width=8.0cm]{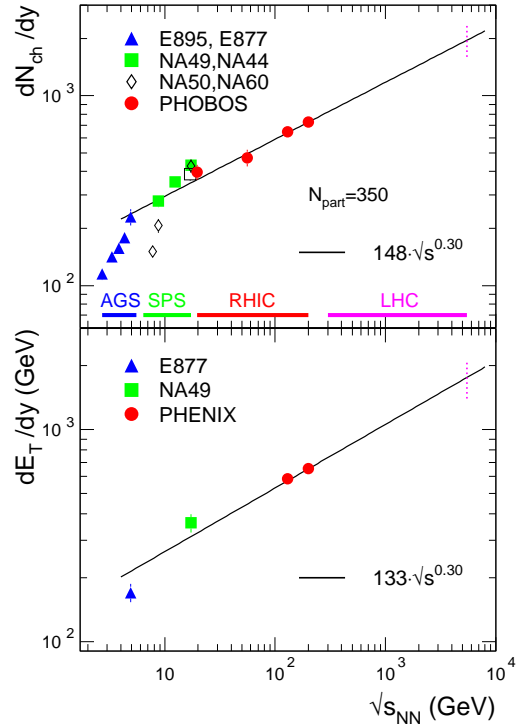}
\vspace{-.80cm}
\caption{Energy dependence of charged particle and transverse energy
  production at midrapidity in central nucleus-nucleus collisions \cite{andronic1}}
\label{dndy}
\end{center}
\end{figure}

In Fig.~\ref{dndy} we show the current information on the energy dependence of
charged particle and transverse energy production near midrapidity in central
PbPb (AuAu) collisions. Overall there is good agreement between experiments. A
noteworthy result of these studies is the apparent power-law increase in both
quantities, close to what Landau \cite{landau} predicted in 1952 in his
hydrodynamic approach. Extrapolation of this trend towards higher energies
would lead to a charged 
particle multiplicity of about 2000 at midrapidity in PbPb collisions at LHC
energy. 

Contrasting these smooth energy dependences are recent observations by the
CERN NA49 collaboration of a rather narrow structure in the K$^+/\pi^+$ ratio
near $\sqrt{s_{nn}} = 8$ GeV \cite{marek}.  These results look rather striking,
but would need to be confirmed by another experiment before further-reaching
conclusions can be drawn.

\begin{figure}[hbt]
\begin{center}
\vspace{-1.0cm}
\includegraphics[width=16.0cm]{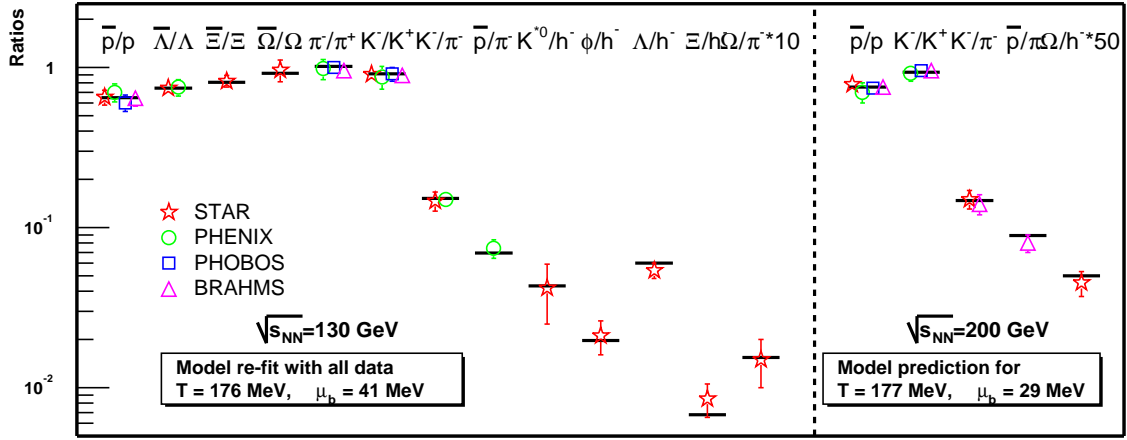} 
\vspace{-.80cm}
\caption{Fit of particle ratios for Au-Au collisions at RHIC.
The measurements are the symbols, the thermal model values are the lines
   \cite{rhic,review}}
\label{yields}
\end{center}
\end{figure}

Turning to more detailed information on hadron production we note that hadron
yields observed in 
central nuclear collisions at AGS, SPS, and RHIC energies are found to be
described with high precision within a hadro-chemical equilibrium
approach \cite{agssps,satz,heppe,cley,beca1,rhic,nu,beca2,rapp}, governed
by a chemical freeze-out temperature T$_{ch}$, baryo-chemical
potential $\mu$ and the fireball volume V$_{ch}$. A recent review can
be found in \cite{review}.  Importantly, the data at SPS and RHIC
energy comprise multi-strange hadrons including the $\Omega$ and $\bar
\Omega$. Their yields agree very well with the chemical equilibrium calculation
and are strongly enhanced as compared to observations in pp
collisions. We show as an example the results for RHIC energy in
Fig.~\ref{yields}.

\begin{figure}[hbt]
\begin{center}
\vspace{-3.0cm}
\includegraphics[width=12.0cm]{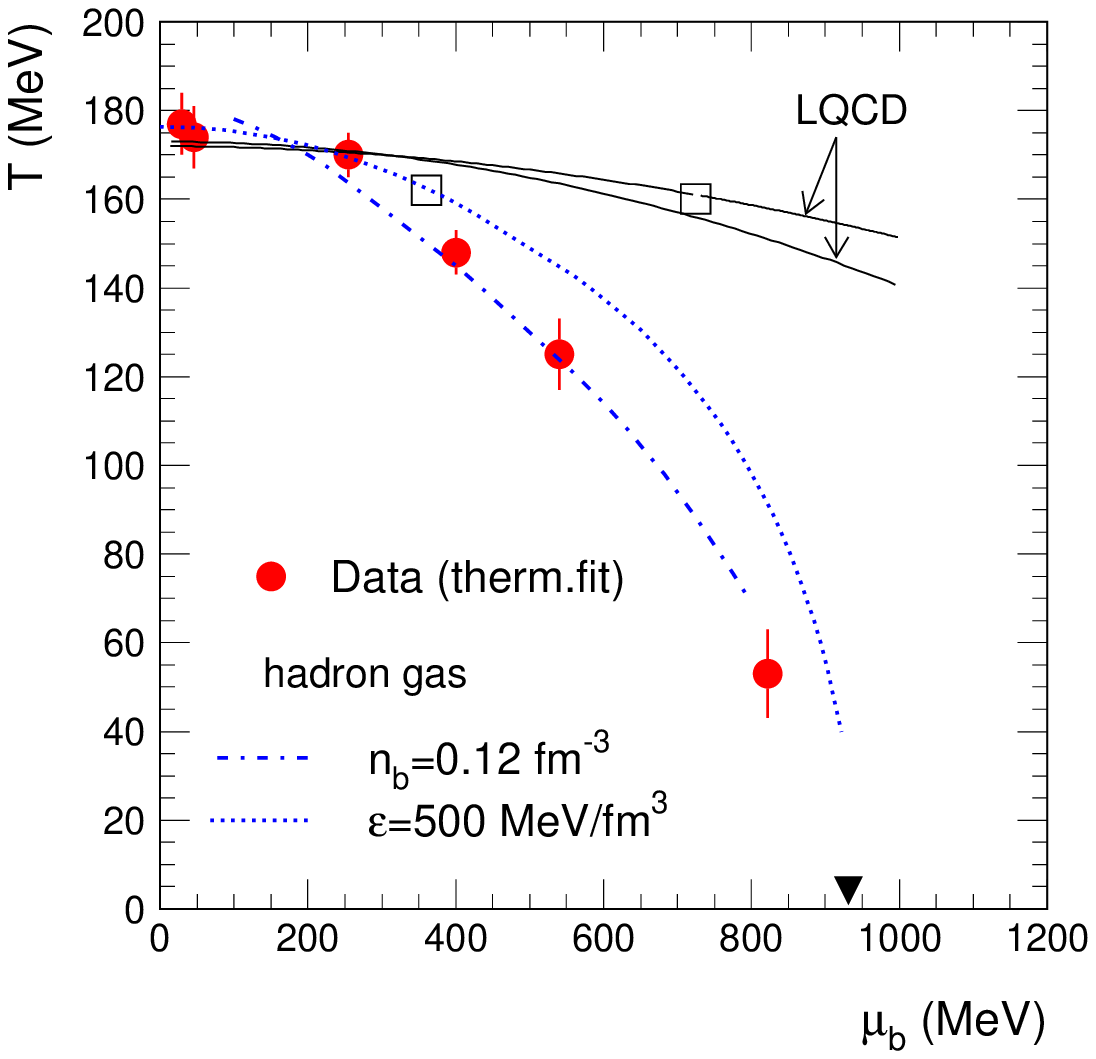} 
\vspace{-.80cm}
\caption{Phase diagram of hadronic matter and chemical freeze-out points
  \cite{andronic1}. The 
  open squares represent recent estimates of the tri-critical point (see text).
}
\label{phase}
\end{center}
\end{figure}

The chemical parameters T$_{ch}$ and $\mu$ determined from fits to data at all
available energies are plotted in the phase diagram of hadronic matter shown
in Fig.~\ref{phase}. The phase transition lines in this figure are taken from
recent analyses within the framework of lattice QCD \cite{karsch2,fodor} and
include the most recent estimate for the tri-critical point. Also included in
this figure is a line of constant energy density, computed within the
framework of the hadronic gas model of \cite{heppe}. Interestingly, this line
of constant energy density is, at $\mu > 500$ MeV, much closer to the chemical
freeze-out points than to the phase lines from lattice QCD. This raises the
question whether a critical energy density significantly exceeding 1
GeV/fm$^3$ as one would extract at critical points such as T$_c = 160$ MeV at
$\mu_c = 700$ MeV makes sense (see Fig.~\ref{phase}).

An important observation about the phase diagram is that, for top SPS energy
and above, the chemical parameters determined from the measured hadron yields
coincide within the uncertainties of about $\pm 10$ MeV with the phase
boundary as determined from lattice QCD calculations.  A natural question
arises: is this coincidence accidental and, if not, what enforces
equilibration at the phase boundary?  Considerations about collisional rates
and timescales of the hadronic fireball expansion \cite{wetterich} imply that
at SPS and RHIC the equilibrium cannot be established in the hadronic medium
and that is is the phase transition which drives the particles densities and
ensures chemical equilibrium.

In  \cite{wetterich} it is further shown that many body collisions near T$_c$
provide  a mechanism for the equilibration.  Because of the rapid density
change near a phase transition such 
multi-particle collisions provide a natural explanation for the observation of
chemical equilibration at RHIC energies and lead to T$_{ch}$=T$_c$ to within
an accuracy of a few MeV. Any scenario with T$_{ch}$ substantially smaller
than T$_c$ would require that either multi-particle interactions dominate even
much below T$_c$ or that the two-particle cross sections are larger than in
the vacuum by a high factor. Both of the latter hypothesis seem unlikely in
view of the rapid density decrease. The critical temperature determined from
RHIC for T$_{ch}\approx \rm{T_c}$ coincides well with lattice estimates
\cite{karsch} for $\mu=0$, as discussed above.  The same arguments as
discussed here for RHIC energy also hold for SPS energies: it is likely that
also there the phase transition drives the particle densities and insures
chemical equilibration.

It was alternatively proposed \cite{stock,heinz2,heinz1} that the observed
hadron 
abundances arise from a direct production of strange (and non-strange)
particles by hadronization. How this happens microscopically is
unclear.  To escape the above  argument that
T$_{ch}$=T$_c$ one would have to argue that no  hadronic
picture for this process exists at all - this is unlikely since the
abundances are determined by hadronic properties (masses) with high
precision. Second, one may question if the ``chemical temperature''
extracted from the abundances is a universal temperature which also
governs the local kinetic aspects and can be associated with the
critical temperature of a phase transition in equilibrium. Indeed, in
a prethermalization process, different equilibrium properties are
realized at different time scales. Nevertheless, all experience shows
that kinetic equilibration occurs before chemical equilibration. It
seems hard to imagine that chemical equilibrium abundances are
realized at a time when the kinematic distributions
are not yet close to their equilibrium
values. 

Furthermore, the recently observed centrality dependence of
multi-strange baryon production at RHIC \cite{star_hyperons} supports the
picture of equilibration near T$_c$ in the hadronic phase: the yields of
hyperons reach the equilibrium value only for rather central collisions
(N$_{part} > 200$), underlining the importance of density in the equilibration
process.

Thermal models have also been successfully used \cite{beca3} to
describe hadron production in e$^+$e$^-$ and hadron-hadron collisions,
leading to temperature parameters close to 170 MeV. Indeed,this
suggests that hadronization itself can be seen as a prethermalization
process. However, to account for the strangeness undersaturation in
such collisions, multi-strange baryons can only be reproduced by
introducing a strangeness suppression factor of about 0.5, leading to
a factor of 8 suppression of $\Omega$ baryons.  In the
hadronic picture this is due to the ''absence'' of sufficient
multi-particle scattering since the system is not in a high density
phase due to a phase transition. We further note, as is evident from
Fig.~\ref{phase} that, in heavy ion collisions, the chemical temperature is
{\bf not} universal, but rather strongly varies (at least at large
$\mu$-values) with the chemical potential, implying the existence of a medium
in distinction to the situation in elementary particle collisions.

\section{Low mass dileptons}

\begin{figure}[hbt]
\begin{center}
\vspace{-0.1cm}
\includegraphics[width=10.0cm]{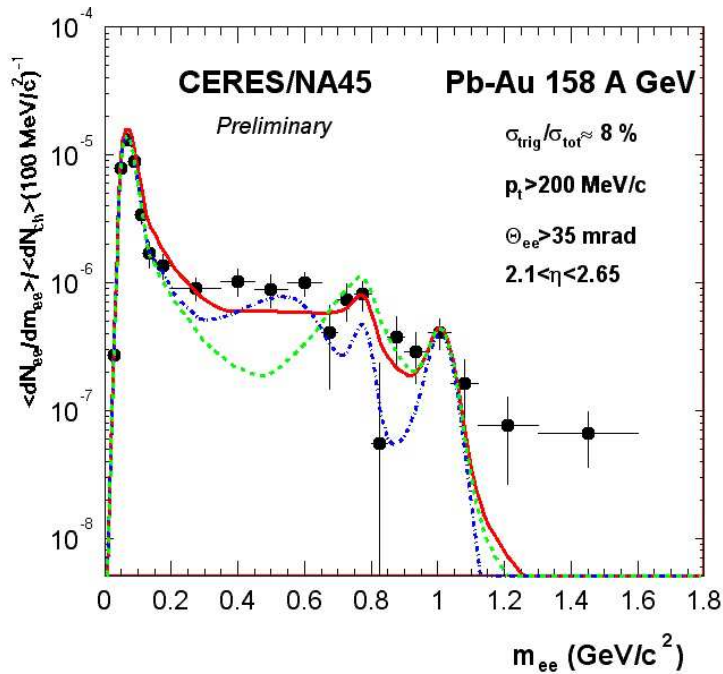} 
\vspace{-.80cm}
\caption{Dielectron mass spectrum from CERES \cite{ceres1} compared to various theoretical
  predictions. For details see text.}
\label{ceres}
\end{center}
\end{figure}

During the year 2000 the CERES experiment took a large data sample of low mass
($ <$ 1.4 GeV) with an apparatus improved by the addition of a radial TPC, for
158 A GeV Pb+Au collisions \cite{ceres1}. Approximately 3$\cdot 10^7$ central
events were recorded with a mass resolution of $\delta m/m \leq 4 \%$. First
results are shown in Fig.~\ref{ceres}. For the first time, structure due to
the $\omega$ and $\phi$ resonances are visible. The data, normalized to the
$\pi^0$ Dalitz decay peak of the hadronic decay cocktail, are compared to
various theoretical predictions. Adding pion annihilation to the hadronic
decay cocktail (dashed line) yields a reasonable description of the $\omega$
and $\phi$ mass region but the continuum in the mass range $0.2 < m_{e^+e^-} <
700 $ MeV is significantly underestimated. Modifying the $\rho$ propagator in
the medium of the fireball according to the scaling proposal of Brown and Rho
\cite{br} (dot-dashed line) or via $\rho$-baryon interactions \cite{rw} (solid
line) leads to much better agreement with the data, strongly disfavoring a
scenario in which the $\rho$ meson is not modified in the hot and dense
medium. We observe that, near the $\omega$ region, the data are better described
in the scenario of \cite{rw} but a stronger statement has to await the final
data analysis.

The $\phi$ meson production yield determined from these data is consistent
with that predicted in the thermal model of \cite{heppe}, but disagrees at the
2 sigma level with the yield determined by the NA49 experiment in the KK
channel \cite{na49phi}. Preliminary results from $\phi \to $KK by
CERES \cite{ceres1} indicate good agreement between the hadronic measurements
where the coverage of the experiments overlap.

\section{High p$_t$ spectra and ``jet'' quenching}
\begin{figure}[hbt]
\begin{center}
\vspace{-0.5cm}
\includegraphics[width=15.5cm]{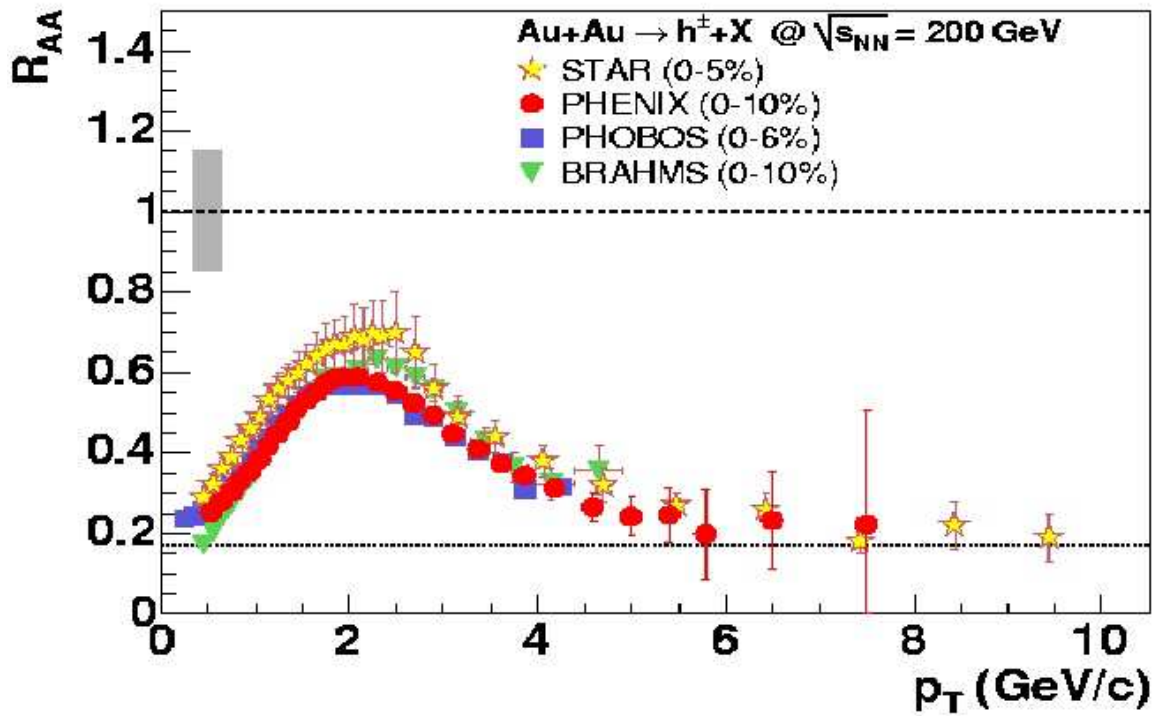} 
\vspace{-.80cm}
\caption{Transverse momentum dependence of the ratio R$_{AA}$ of charged
  particle yields in central AA and pp collisions. For details see text.}
\label{raa1}
\end{center}
\end{figure}

One of the major results from all 4 RHIC experiments is the observation, at
p$_t$ values larger than 2 GeV, of a strong suppression of charged particle
production in central nucleus-nucleus (AA) collisions compared to pp
collisions. Commonly this is expressed in the ratio R$_{AA}=
\frac{d^2N^{AA}/dp_td\eta}{\langle N_{binary}\rangle d^2N^{pp}/dp_td\eta}$,
i.e. the ratio of the spectra scaled by the number of binary collisions.  In
the low p$_t$ region where participant scaling applies, R$_{AA} \approx
N_{part}/\langle N_{binary}\rangle \approx 0.45$ for central PbPb
collisions. In the hard scattering region, i.e.in the p$_t$ region above a few
GeV this ratio should approach 1 if a nuclear collision is merely a
superposition of binary nucleon-nucleon collisions. In contradistinction to
this expectation, R$_{AA}$ for central AuAu collisions never exceeds 0.7, and
quite dramatically drops to a value near 0.2 for p$_t >$ 6 GeV and levels off
there \footnote{For a recent review and original references see M. Gyulassy
and L. McLerran \cite{gyumcl}.}.  This is demonstrated in Fig.~\ref{raa1},
where the impressive consistency between the data from all 4 RHIC experiments
is also apparent. No or little deviation from binary scaling is observed
\cite{d_au} in d Au collisions, demonstrating that the suppression is due to
the medium produced in AA collisions and not an initial state effect. The most
straightforward interpretation of this dramatic effect is that the parton
preceding the high p$_t$ hadron observed in the measurements scatters and
loses energy in the medium produced in the AA collision. The degree of opacity
of the medium can be evaluated more quantitatively: for a completely black
medium one would expect only surface emission of hadrons in the relevant p$_t$
range of 3 - 10 GeV.  This implies N$_{part}^{2/3}$ scaling, leading to a
lower limit of 
\begin{equation}
 R^{min}_{AA} =\frac{2^{4/3}}{\rm{N}_{part}^{2/3}} \approx 0.05
\end{equation}
for central AuAu collisions, i.e. a factor of 4 lower than observed at
RHIC. It will be interesting to find out whether this surface emission limit
will be reached at LHC energies.

Recent measurements by the BRAHMS collaboration \cite{brahms_forward} for dAu
collisions show that R$_{dAu}$ decreases to values significantly smaller than
1 with increasing hadron rapidity.  This has been interpreted \cite{gyumcl}
as evidence for non-linear effects expected within the framework of the
color-glass condensate (CGC). However, recent analysis \cite{guzey} shows that
even in the Brahms experiment only rather moderate parton momentum fractions
x$> 0.01$ contribute significantly to the cross section. Further work needs to
be done to substantiate the CGC interpretation.

It is important to realize that the overall suppression visible in
Fig.~\ref{raa1} for charged hadrons depends in detail significantly on the
type of hadron under consideration. In particular, all mesons seem to follow a
similar suppression pattern while the baryon/meson ratio increases strongly
with transverse momentum, reaching a value close to 1 for the p/$\pi^+$ ratio,
e.g., near p$_t = 3$ GeV \cite{ph1,st1}.  This behavior is quite inconsistent
with the standard fragmentation picture of a hadronizing parton, from which
much smaller p/$\pi^+$ ratios are expected. Based on these and other data on
elliptic flow of mesons and baryons in the transverse momentum region $2 <
\rm{p_t} < 5$ GeV an alternative picture was developed recently
\cite{voloshin,fries,greco,hwa} in which hadron production occurs by
recombination rather than fragmentation of partons. In fact, if the parton
density is high enough, there is, for an exponentially decreasing parton p$_t$
spectrum, a critical parton transverse momentum below which
``naive''\footnote{Naive recombination implies that 2 partons at
$\vec{\rm{p_{1}}}$ and $\vec{\rm{p_{2}}}$ will recombine to form a hadron at
$\vec{\rm{p_{12}}}$. Note, however, that this process violates entropy
conservation.}  recombination will win over fragmentation. First calculations
within this framework indicate that the enhanced baryon emission as well as
the hadron species dependence of the elliptic flow pattern can be quite well
accounted for. How to reconcile this picture which is based on an expanding
parton phase with the observed jet-like correlations in the data seems not
straightforward (see, e.g., the discussion in \cite{fries2}).

\begin{figure}[htb]
\begin{tabular}{ccc}
\begin{minipage}{.43\textwidth}
\centering\includegraphics[width=1.03\textwidth,height=1.14\textwidth]{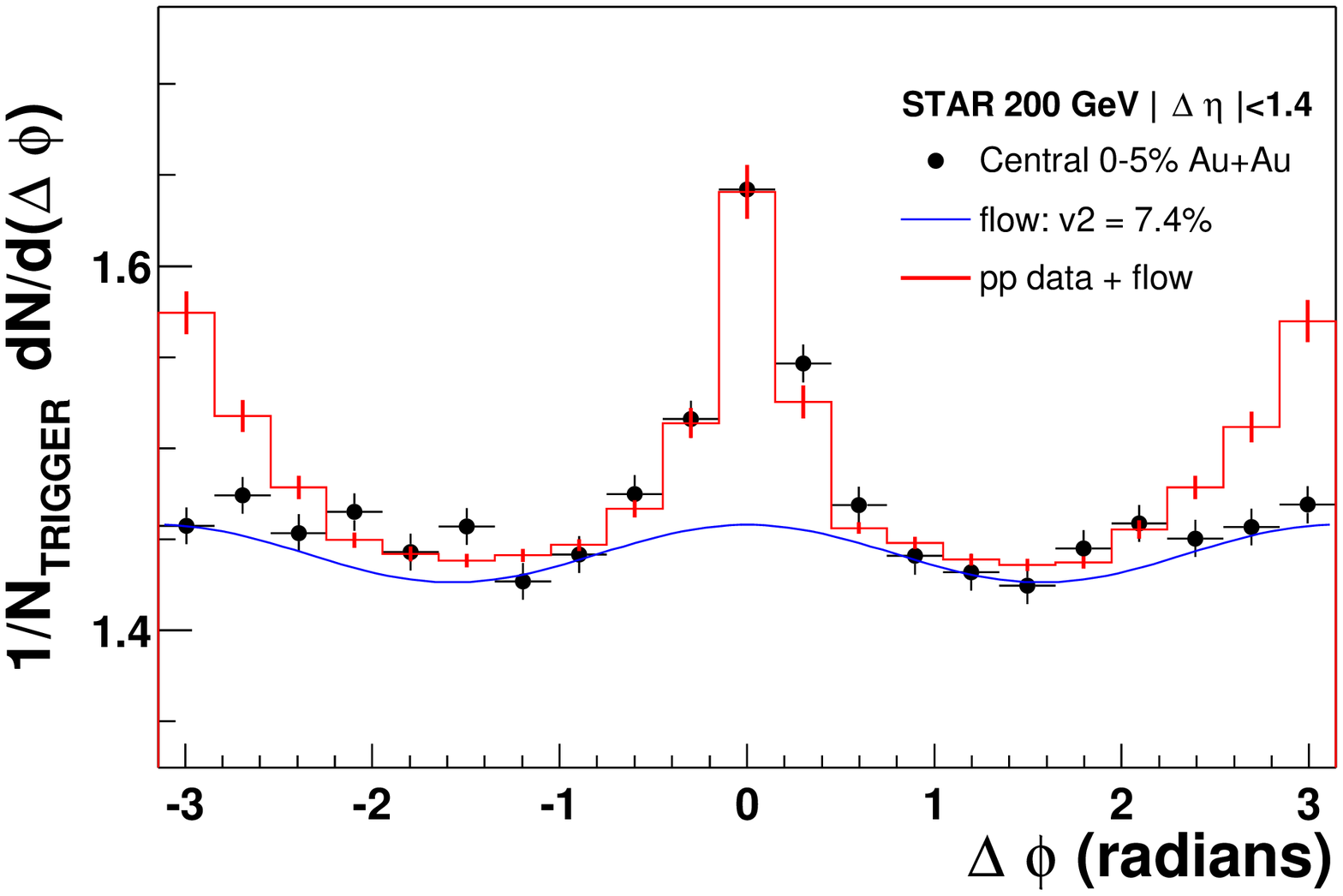}
\end{minipage}  & \begin{minipage}{.43\textwidth}
\centering\includegraphics[width=1.03\textwidth,height=1.14\textwidth]{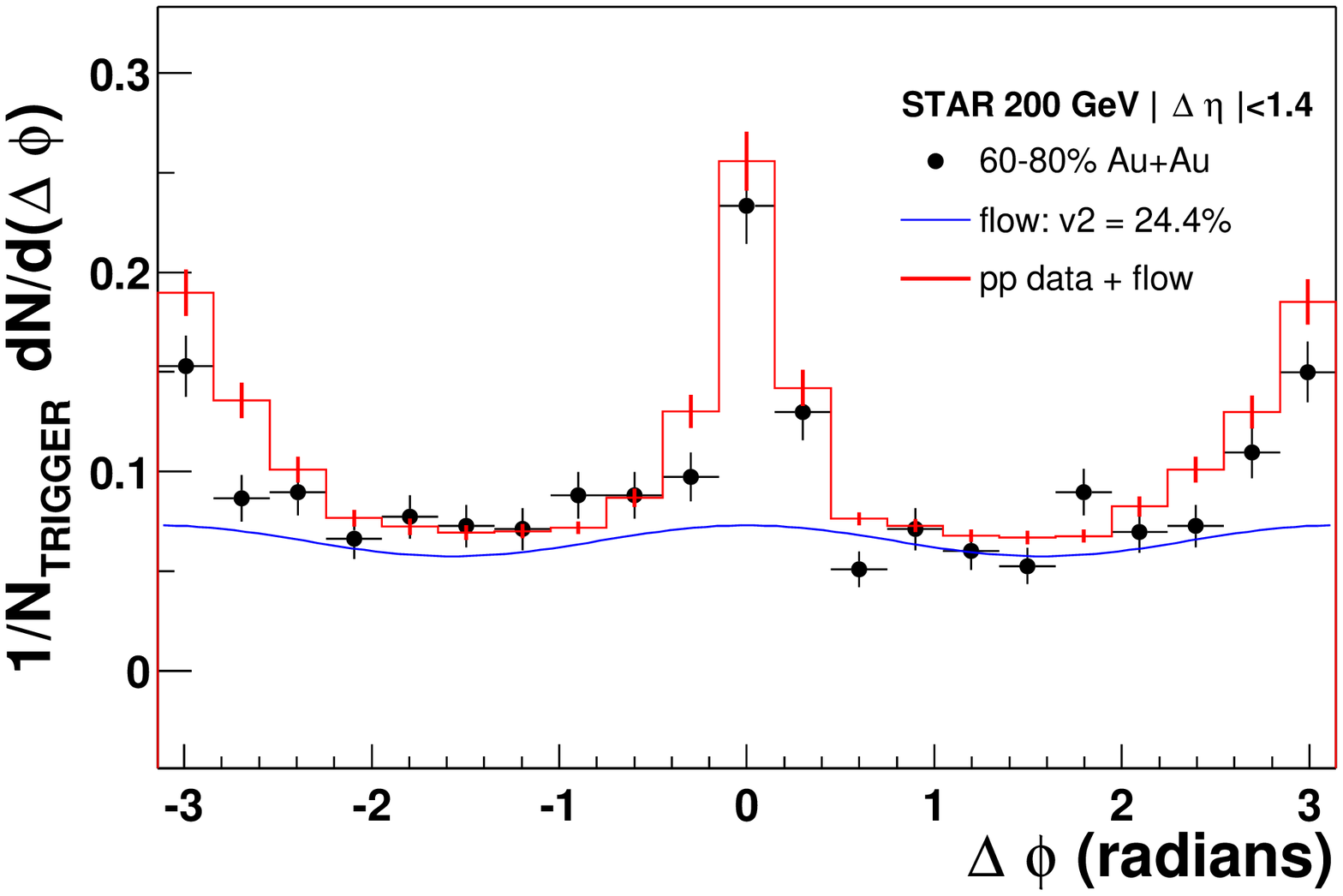}
\end{minipage}
\end{tabular}
\caption{Azimuthal correlations in pp and peripheral Au-Au collisions (right
  panel) and in central Au-Au collisions (left panel), where the dijet
  pattern is absent.} 
\label{correl}
\end{figure}

Further evidence for jets and their interaction with the dense medium formed
in the collision comes from measurements of azimuthal angular correlations.
Results by the STAR collaboration \cite{jet1,jet2} of such correlation
measurements in pp and 
AuAu collisions are shown in Fig.~\ref{correl}. In these measurements a
leading hadron with p$_t > 4$ GeV defines the ``near side'' direction. The
azimuthal correlation of all charged particles with p$_t > 2$ GeV is depicted
in this figure for pp and peripheral AuAu collisions where a typical jet-like
structure is observed with pronounced peaks centered at $\Delta\phi = 0$ and
$\pi$. Note, however, the striking absence of the away-side ($\Delta\phi =
\pi$) peak in central AuAu collisions, implying that the medium formed in such
collisions either widens this peak dramatically or shifts the momenta of
particles in the away-side direction below the 2 GeV threshold. Evidence for 
such a widening was observed by the CERES collaboration in PbAu collisions at
SPS energy \cite{ceres_corr}, indicating that the onset of this jet-quenching
phenomenon takes place already at much lower energy.  The data presented in
Fig.~\ref{correl} clearly suggest that two-body parton-parton collisions and
the modification of parton momenta by the medium are at the heart of the
observed phenomena.

Such energy loss of partons or ``quenching'' of the resulting jets was
predicted as due to  induced gluon radiative energy loss in the dense medium
\cite{gyu_plu} \footnote{Energy loss by elastic parton rescattering in the
  dense medium  as
  originally proposed by Bjorken \cite{bj} is now understood to be of order 1
  GeV/fm, i.e. comparable to the string tension or the energy loss in cold
  nuclear matter, much too small to explain jet quenching.}. The theory of
parton energy loss in a dense gluonic medium was since significantly improved
\cite{baier,baier_r} to take fully account of the coherent part of the gluon
radiative spectrum which is induced by multiple collisions in the medium. This
leads to an energy loss larger by approximately an order of magnitude compared
to that induced by elastic scattering.

\begin{figure}[htb]
\begin{tabular}{ccc}
\begin{minipage}{.43\textwidth}
\centering\includegraphics[width=1.03\textwidth,height=1.14\textwidth]{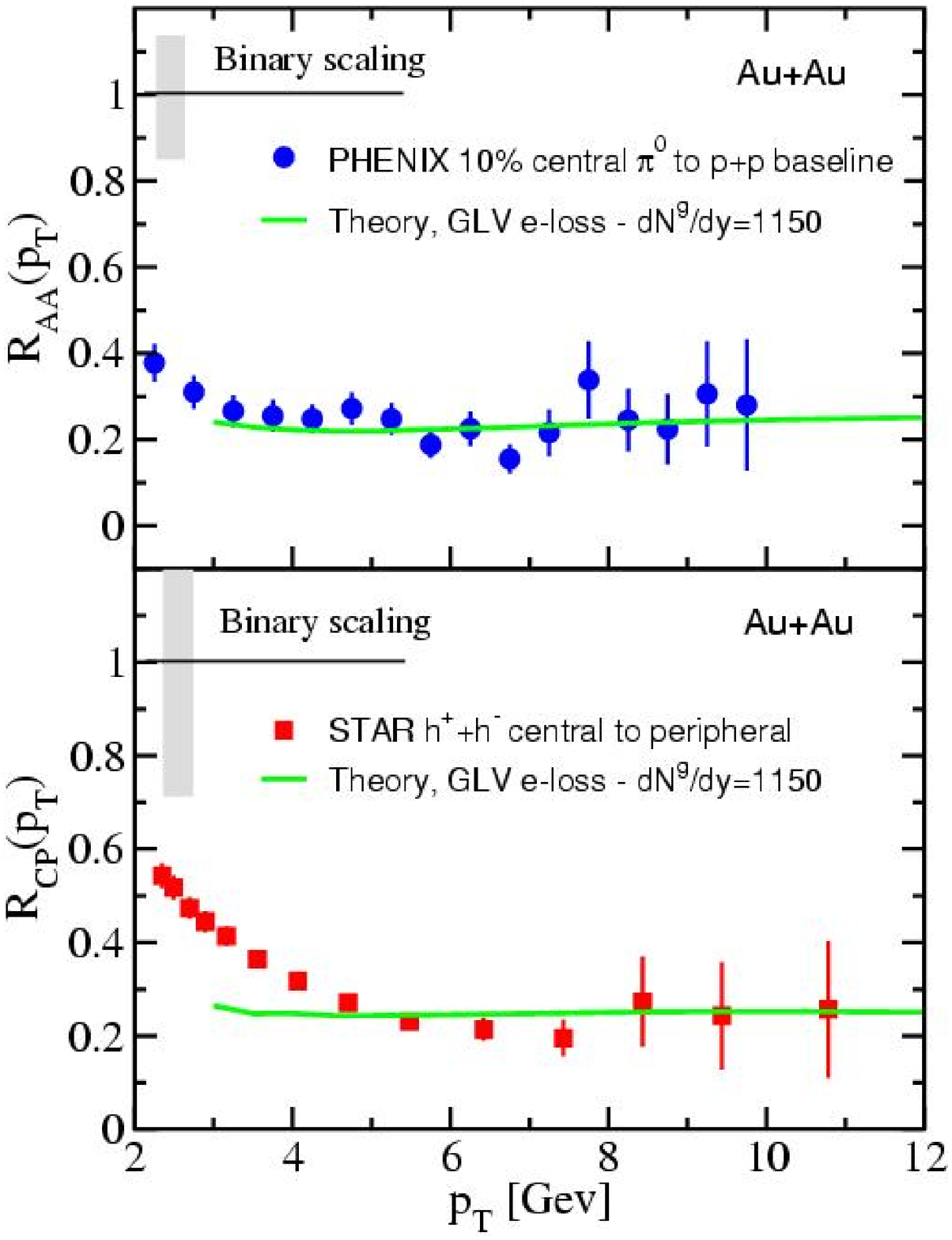}
\end{minipage}  & \begin{minipage}{.43\textwidth}
\centering\includegraphics[width=1.03\textwidth,height=1.14\textwidth]{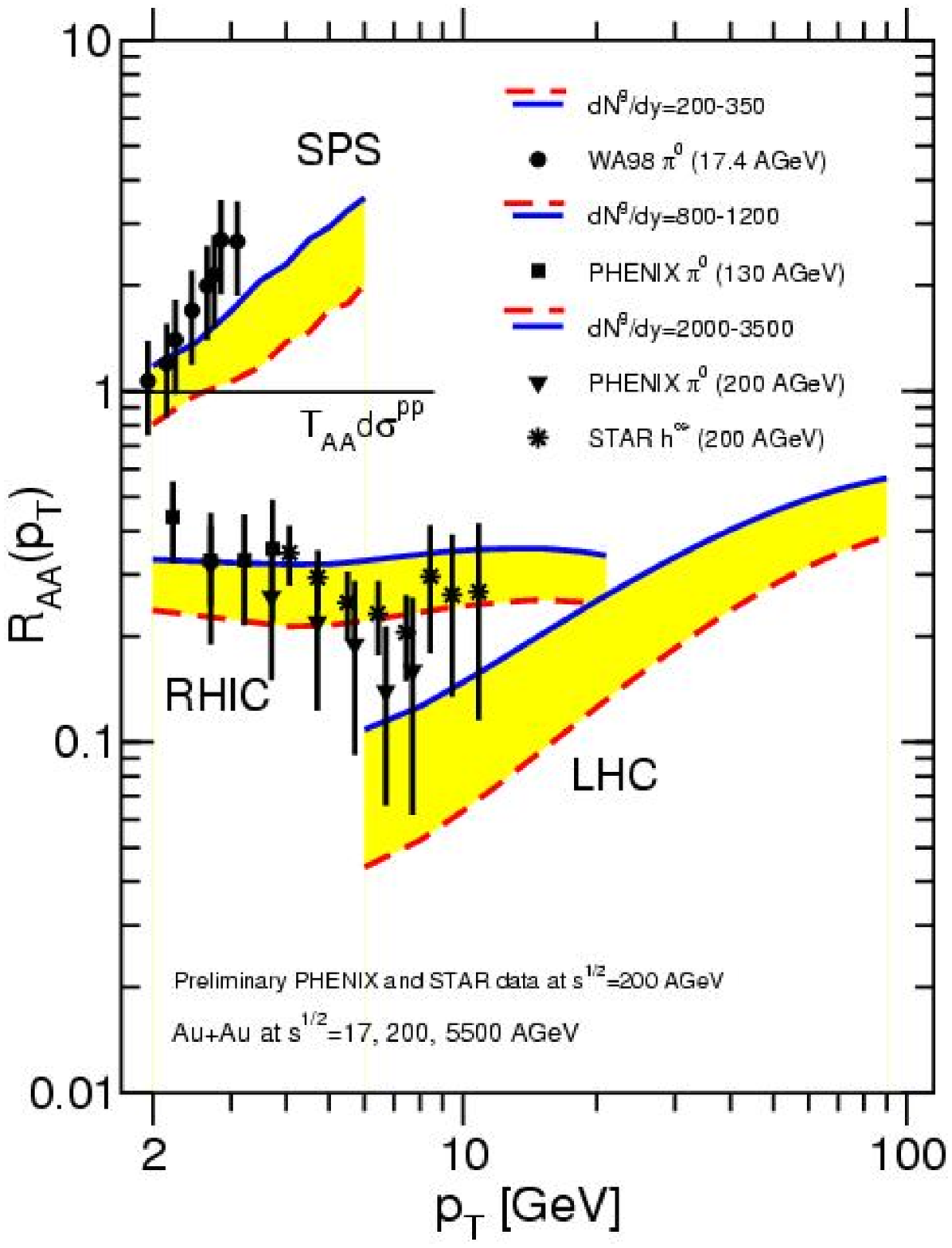}
\end{minipage}
\end{tabular}
\caption{Comparison of energy loss calculations based on the GLV formalism
  \cite{glv,vitev} with RHIC data and predictions for LHC energy.}
\label{vitevcp}
\end{figure}

Based on these developments first
theoretical interpretations of the observed quenching effect have appeared. As
an example we present in Fig.~\ref{vitevcp} a recent comparison to RHIC data
and predictions for LHC energy. The comparison of these calculations to the
RHIC data shows that the experimental observations are consistent with an
initial effective gluon rapidity density $\frac{dN_g}{dy} \approx 1000$ at a
mean transverse momentum of p$_t \approx 1$ GeV. 

Using this result one may estimate the initial energy density at time $\tau =
1/\rm{p_t}$ to be $\epsilon_i =\frac{p_t^2}{\pi R^2}*\frac{dN_g}{dy} \approx
33$ GeV/fm$^3$.  Clearly this estimate exceeds by 2 orders of magnitude the
energy density inside a heavy nucleus and is more than a factor 30 above the
critical energy density for the transition from hadronic matter to the
QGP. The jet quenching data thus lend strong support to the interpretation
that a quasi-equilibrated partonic state is formed at early times in central
ultra-relativistic nuclear collisions.

\section{Summary and outlook}

In this brief survey we have presented strong evidence for the formation of
partonic matter near equilibrium from data on hadron production at low and
intermediate transverse momenta in ultra-relativistic nuclear collisions. In
particular, the observation that the data on hadron production ratios are in
very good agreement with chemical equilibrium predictions has been used to
provide a first determination of the critical temperature of the QCD phase
transition. The striking data on jet quenching demonstrate the opacity of the
new state of matter, and provide a glimpse into the very hot early phase of
the partonic fireball. 
First and tantalizing data on heavy quark and quarkonia production have
already been obtained at RHIC and high precision data in this area are eagerly
awaited. Also the SPS program on  heavy quarks will be revived with new results
from the NA60 experiment, which will furthermore provide high resolution, high
statistics pA and InIn data on low-mass dilepton production. 
With LHC coming on-line in 2007 quark matter studies will continue on
much hotter and longer-lived fireballs formed at $\sqrt{s_{nn}} = 5.5$
TeV. For a survey of the planned physics program see, e.g.,
\cite{schutz,alice_ppr}. Matter at the highest baryon density will be studied
with precision experiments at the planned FAIR facility at GSI. For a survey
of the planned physics program see, e.g., \cite{senger}.  

\section{Acknowledgement} 
I would like to thank Dr. D. Miskowiec for a critical reading of the manuscript.


\begin{thebibliography}{9}
\bibitem{hagedorn} R. Hagedorn, Nuovo Cimento Suppl. {\bf 3} (1965) 147 
\bibitem{wilczek} D.J. Gross, F. Wilczek, Phys. Rev. Lett. {30} (1973) 1343.
\bibitem{politzer} H.D. Politzer, Phys. Rev. Lett. {30} (1973) 1346.
\bibitem{collins} J.C. Collins and M.J. Perry, Phys. Rev. Lett. {\bf 34}
  (1975) 1353. 
\bibitem{cabibbo} N. Cabibbo and G. Parisi, Phys. Lett. {\bf B59} (1975) 67.
\bibitem{hagedorn1} R. Hagedorn, CERN-TH-3918/84.
\bibitem{baym} G. Baym, Nucl. Phys. {\bf A698 } (2002) xxxii.
\bibitem{karsch} F. Karsch, E. Laermann, A. Peikert, Nucl. Phys. {\bf A605}
  (2001) 579.
\bibitem{karsch1} S. Ejiri et al, hep-lat/0408046.
\bibitem{nantes} Proc. Quark Matter 2002, Nucl. Phys. {\bf A715} (2003),
  H. Gutbrod, J. Aichelin, and K. Werner, editors. 
\bibitem{oakland} Proc. Quark Matter 2004, J. Phys. G: Nucl. Part. Phys. {\bf
  30} (2004), H. G. Ritter and X.N. Wang, editors.
\bibitem{andronic1} A. Andronic and P. Braun-Munzinger, hep-ph/0402291.
\bibitem{landau} L.D. Landau, Izv. Akad. Nauk Ser. Fiz. {\bf 5} (1950) 570.
\bibitem{marek} M. Gazdzicki et al, NA49 collaboration, QM2004. 
\bibitem{agssps} P. Braun-Munzinger,
  J.  Stachel, J.P. Wessels and   N. Xu, Phys. Lett. {\bf B344} (1995) 43 and
  {\bf B365} (1996) 1. 
\bibitem{satz} J. Cleymans, D. Elliott, H. Satz, and R.L. Thews, Z. Phys. {\bf
  C74} (1997) 319.
\bibitem{heppe}P. Braun-Munzinger, I. Heppe and J. Stachel, Phys. Lett. {\bf
B465} (1999) 15.
\bibitem{cley}J. Cleymans and K. Redlich, Phys. Rev. {\bf C60} (1999) 054908.
\bibitem{beca1} F. Becattini, J. Cleymans, A. Keranen, E. Suhonen, and
K. Redlich, Phys. Rev. {\bf C64} (2001) 024901.
\bibitem{rhic} P. Braun-Munzinger, D. Magestro, K. Redlich, and
  J. Stachel, Phys. Lett. {\bf B518} (2001) 41.
\bibitem{nu}N. Xu and M. Kaneta, Nucl. Phys. {\bf A698} (2002) 306c.
\bibitem{beca2} F. Becattini, J. Phys. {\bf G28} (2002) 1553.
\bibitem{rapp} R. Rapp and E. Shuryak, Phys. Rev. Lett. {\bf 86} (2001) 2980.
\bibitem{review}P. Braun-Munzinger, K. Redlich, and J. Stachel,
  nucl-th/0304013, invited review in Quark Gluon Plasma 3, eds. R.C. Hwa and
  X.N. Wang, (World Scientific Publishing, 2004).
\bibitem{karsch2} S. Ejiri et al., hep-lat/0312006.
\bibitem{fodor} Z. Fodor and S. Katz, JHEP {\bf 0404} (2004) 050.
\bibitem{wetterich} P. Braun-Munzinger, J. Stachel, C. Wetterich,
  Phys. Lett. {\bf B596} (2004) 61.
\bibitem{stock} R. Stock, Phys. Lett. {\bf B465} (1999) 277.

\bibitem{heinz1} U. Heinz, Nucl. Phys. {\bf A638} (1998) 357.
\bibitem{heinz2} U. Heinz, Nucl. Phys. {\bf A685} (2001) 414.
\bibitem{star_hyperons} J. Adams et al., STAR coll., Phys. Rev. Lett. {\bf 92}
  (2004) 182301, nucl-ex/0307024.
\bibitem{beca3}  F. Becattini, Z. Physik {\bf C69} (1996) 485,
F. Becattini and U. Heinz, Z. Physik {\bf C76} (1997) 269.
\bibitem{ceres1} A. Marin et al., CERES coll., J. Phys. {\bf G30} (2004) S709.
\bibitem{br} G.E. Brown and M. Rho, Phys. Rep. {\bf 363} (2002) 85.
\bibitem{rw} R. Rapp and J. Wambach, Adv. Nucl. Phys. {\bf 25} (2000) 1, and
  R. Rapp, private communication.
\bibitem{na49phi} D. R{\"o}hrich, J. Phys.  {\bf G27} (2001) 355. 
\bibitem{gyumcl} M. Gyulassy, L. McLerran, nucl-th/0405013.
\bibitem{d_au} J. Adams et al., STAR coll., Phys.Rev.Lett. {\bf 91} (2003)
  072304, S. Adler et al., PHENIX coll., Phys.Rev.Lett. {\bf 91} (2003)
  072303; I. Arsene et al., BRAHMS coll., Phys.Rev.Lett. {\bf 91} (2003)
  072305;   B. Back et al., PHOBOS coll., Phys.Rev.Lett. {\bf 91} (2003)
  072302. 
\bibitem{brahms_forward} I. Arsene et al., BRAHMS collaboration, nucl-ex/0403005.
\bibitem{guzey} V. Guzey, M. Strikman, W. Vogelsang, hep-ph/0407201.
\bibitem{ph1} K. Adcox et al., PHENIX coll., Phys. Rev. Lett. {\bf 88} (2002)
  242301. 
\bibitem{st1} C. Adler et al., STAR coll., Phys. Rev. Lett. {\bf 89} (2002)
  093201.
\bibitem{voloshin} D. Molnar and S.A. Voloshin, Phys. Rev. Lett. {\bf 89}
  (2002) 092301.
\bibitem{fries} R. J. Fries, B. M{\"u}ller,, C. Nonaka, S.A. Bass,
  Phys. Rev. Lett. {\bf 90} (2003) 202303
\bibitem{greco} V. Greco, C. M. Ko, P. Levai, Phys. Rev. Lett. {\bf 90} (2003)
  202302.
\bibitem{hwa} R. C. Hwa, C. B. Yang, Phys. Rev. {\bf C67} (2003) 034902.  
\bibitem{fries2} R. J. Fries, S. A. Bass, B. M{\"u}ller, nucl-th/0407102.
\bibitem{jet1} C. Adler et al, STAR coll., Phys. Rev. Lett. {\bf 90}
  (2003) 082302.
\bibitem{jet2} D. Hartke et al, Star coll., Nucl. Phys. {\bf A715} (2003) 272.   
\bibitem{ceres_corr} G. Agakichiev et al., CERES coll., Phys. Rev. Lett. {\bf
  92} (2004) 032301.
\bibitem{gyu_plu} M. Gyulassy, M. Plumer, Nucl. Phys. {\bf A527} (1991) 641.
\bibitem{bj} J. D. Bjorken, Fermilab Pub-82-059-THY (unpublished). 
\bibitem{baier} R. Baier, Yu. Dokshitzer, A. H. Muller, S. Peigne, D. Schiff,
  Nucl. Phys. {\bf B484} (1997) 265.
\bibitem{baier_r} for a recent review see  R. Baier, D. Schiff, B. G. Zakharov,
  Ann. Rev. Nucl. Part. Sci. {\bf 50} (2000) 37 for a recent review.
\bibitem{glv} M. Gyulassy, P. Levai, I. Vitev, Nucl. Phys. {\bf B549} (2001)
  371.
\bibitem{vitev} I. Vitev, J. Phys. {\bf G30} (2004) S791, hep-ph/0403089.
\bibitem{schutz} I. Schutz, J. Phys. {\bf G30} (2004) S903.
\bibitem{alice_ppr} J.P. Revol et al., ALICE physics performance report,
  J. Phys. {\bf G30} (2004) 1517.
\bibitem{senger} P. Senger,J. Phys. {\bf G30} (2004) S1087. 
\end{thebibliography}
\end{document}